\documentclass[preprint,authoryear,11pt]{elsarticle}
\usepackage[margin=0.80in]{geometry}

\usepackage{ccaption}
\usepackage{amsmath}
\usepackage{amsbsy}
\usepackage{dcolumn}
\usepackage{multirow}
\usepackage[table]{xcolor}
\usepackage{wrapfig}
\usepackage{pifont}
\usepackage{float}
\usepackage{adjustbox}
\usepackage{wrapfig}
\usepackage{soul}
\usepackage{footnote}
\usepackage{times}
\usepackage{lipsum}
\usepackage{lineno}
\usepackage[T1]{fontenc}
\usepackage{textcomp}
\usepackage{graphicx}
\usepackage{graphics}
\usepackage{soul}

\usepackage{epstopdf}
\renewcommand{\figurename}{Fig.}
\usepackage{caption}
\usepackage{subcaption}
\usepackage{natbib}
\usepackage{amssymb,amstext,amsmath}
\usepackage{float}
\usepackage{booktabs}
\usepackage{array,ragged2e}
\usepackage[table]{xcolor}
\usepackage{multirow}
\usepackage{algorithm}
\usepackage{algpseudocode}
\usepackage{pifont}
\usepackage{enumerate}
\usepackage[utf8x]{inputenc}

\biboptions{authoryear}
\makeatletter
\def\ps@pprintTitle{%
 \let\@oddhead\@empty
 \let\@evenhead\@empty
 \def\@oddfoot{\centerline{\thepage}}%
 \let\@evenfoot\@oddfoot}
\makeatother
\newtheorem{theorem}{Theorem}
\begin{document}

\begin{frontmatter}

\title{A Combinatorial Approach for Nonparametric Short-Term Estimation of Queue Lengths using Probe Vehicles}

\author{Gurcan Comert\corref{mycorrespondingauthor}}
\cortext[mycorrespondingauthor]{Corresponding author}
\ead{gurcan.comert@benedict.edu}
\address{Computer Science, Physics, and Engineering Department, Benedict College, 1600 Harden St., Columbia, SC USA 29204}
\address{Information Trust Institute, University of Illinois Urbana-Champaign, 1308 West Main St., Urbana, IL 61801 USA}

\author{Tewodros Amdeberhan}
\ead{tamdeber@tulane.edu}
\address{Department of Mathematics, Tulane University, 6823 St. Charles Avenue
New Orleans, LA 70118 USA}

\author{Negash Begashaw}
\ead{negash.begashaw@benedict.edu}
\address{Computer Science, Physics, and Engineering Department, Benedict College, 1600 Harden St., Columbia, SC 29204 USA}

\author{Mashrur Chowdhury}
\ead{mac@clemson.edu}
\address{Glenn Department of Civil Engineering, Clemson University, Lowry Hall, Clemson, SC 29634 USA}
%\vspace{-50mm}
\begin{abstract}
%% ***   Put your Abstract here.   ***
%% (At most 250 words.)
Traffic state estimation plays an important role in facilitating effective traffic management. This study develops a combinatorial approach for nonparametric short-term queue length estimation in terms of cycle-by-cycle partially observed queues from probe vehicles. The method does not assume random arrivals and does not assume any primary parameters or estimation of any parameters but uses simple algebraic expressions that only depend on signal timing. For an approach lane at a traffic intersection, the conditional queue lengths given probe vehicle location, count, time, and analysis interval (e.g., at the end of red signal phase) are represented by a Negative Hypergeometric distribution. The estimators obtained are compared with parametric methods and simple highway capacity manual methods using field test data involving probe vehicles. The analysis indicates that the nonparametric methods presented in this paper match the accuracy of parametric methods used in the field test data for estimating queue lengths. 
\end{abstract}
\end{frontmatter}
%\IEEEpeerreviewmaketitle
\section{Introduction}
Probe vehicles or Lagrangian sensors (\cite{herrera2010incorporation}) can be considered as tracking-device equipped vehicles that can report critical information such as direct travel time, speed (\cite{ramezani2012estimation, jenelius2013travel,jenelius2015probe,hans2015applying,zheng2018traffic}) flow (\cite{duret2017traffic,seo2019fundamental}) or inferred delay (\cite{florin2020towards}) and queue lengths (\cite{bae2019spatio, wang2020queue}) as they traverse transportation networks. Commercial taxis, volunteers, transit buses, maintenance vehicles, commercial trucks, etc., can report their location and timestamps through cellphones and GPS devices for improved traffic operations or better planning. Traffic state can be estimated using Collected data. The accuracy of these estimates depends on the accuracy of reported sensor data and the penetration of the number of data received from the vehicles. Regardless, observing mobile data from transportation networks gives critical coverage for dynamic traffic behavior. This study presents a method for estimating queue length  given that (1) probe vehicles can be observed on a lane accurately and infer the order of vehicles in a queue, (2) we can deduce the beginning of queue start time to probe vehicle arrival times (e.g., relative to the beginning of red duration at a signal), and (3) we can track the number of probe vehicles in the queue. Assuming these data are available, using the combinatorics approach, we develop queue length estimators that can be used for any queues without requiring primary arrival rate or probe vehicle market penetration rate (or percentage) parameters.  

Researchers have extensively studied the queue length estimation problem by proposing parametric (\cite{zhao2019various, zhao2021maximum, zhao2021hidden}) and nonparametric methods. In this paper, we focused on the review of nonparametric ones. Jin and Ma presented a study on a nonparametric Bayesian method for traffic state estimation (\cite{jin2019non}). In their study, they developed a generalized modeling framework for estimating traffic states at signalized intersections. The framework is nonparametric and data-driven, and no explicit traffic flow modeling is required.
%Their study developed a nonparametric data-driven framework for traffic state estimation at signalized intersections. It does not require explicit modeling of traffic dynamics. It essentially applies a Bayesian filter and recursively calculates the traffic parameter estimates.
Wong et al. estimated the market penetration rate (probe proportion or percentage) (\cite{wong2019estimation}). Based on probe vehicle data alone, they proposed a simple, analytical, nonparametric, and unbiased approach to estimate penetration rate.
%They proposed an analytical and nonparametric estimation method for probe vehicle (PV) market penetration rate. 
The method fuses two estimation methods. One is from probabilistic estimation and the second from samples of probe vehicles which is not affected by arrival patterns. It uses PVs and all vehicles ahead of the last PV in the queue.

Gao et al. presented queue length estimations (QLEs) based on shockwaves and backpropagation neural network (NN) sensing (\cite{gao2019connected}). The approach uses PV data and queue formation dynamics. It uses the shockwave velocity to predict the queue length of the non-probe vehicles. The NN is trained with historical PV data. The queue lengths at the intersection are obtained by combining the shockwave and NN-based estimates by variable weight. Tan et al. introduced License Plate Recognition (LPR) data in their study to fuse with the vehicle trajectory data, and then developed a lane-based queue length estimation method (\cite{tan2020fuzing}). Authors matched the LPR with probe vehicle data. They obtained the probability density function of the discharge headway and the stop-line crossing time of vehicles. They presented the lane-based queue lengths and overflow queues. Wang et al. proposed a QLE method on street networks using occupancy data (\cite{wang2013modeling}). Their key idea is using the speed decrease as the queue increase downstream of loop detector. This would result in higher occupancy at constant volume-to-capacity ratios. 
%Then, authors generated data for various link length, lane width, lane number, and bus ratio using VISSIM simulation. 
Using VISSM simulation, they generated data for various link length, lane number, and bus ratio.
They fit a logistic model for the queue length and occupancy relationship. Then, queue lengths were estimated using multiple regression models.

%In this paper, we aim to generally model queue lengths at intersections without assuming random arrivals or any primary parameters or estimating parameters.
The purpose of this paper is to model queue lengths at intersections generally without assuming random arrivals or any primary parameters or estimating parameters. 
Unlike fundamental non-parametric queue length estimations from arrival and service distributions (\cite{schweer2015nonparametric, goldenshluger2016nonparametric, goldenshluger2019nonparametric,singh2021estimation}), our method  uses mathematical techniques from combinatorics to derive discrete conditional probability mass functions of observed information about the queue and derive moments of the distributions without depending on arrival or service distributions. The approach presented in this study essentially extends the results from (\cite{Comert2009, Comert2013}) where \cite{Comert2009} presented a conditional probability mass function for probe location information and \cite{Comert2013} provided closed-form queue length estimators given probe vehicle location and time information for Poisson arrivals.     
%The paper is organized as follows. 

The paper is organized as follows: In section \ref{intro}, the approach is defined to set up derivations. In section \ref{PEV}, we use combinatorial arguments and present a closed form of the sum of the probabilities in Eqs.~(\ref{probability-1}) and (\ref{probability-2}). The result obtained in section \ref{PEV} enables us to define a probability mass function. We show that this probability distribution is Negative Hypergeometric. We use the results for the mean and variance of the distribution to derive formulas for the queue length estimators. In section \ref{field}, we present numerical examples for the behavior of the derived estimators and show the performance of the estimators using field data. 
%Finally, in section \ref{sctconc} we summarize our findings and discuss possible future research directions.
We summarize our findings and discuss possible future research directions in section \ref{sctconc}.
\section{Problem Definition}
\label{intro}
%CV environment-V2I environment-why cyber detection is important and why we need real-time detection
%explain different algorithms and why we choose EM and CUSUM for real time detection
Probe vehicles (PVs) and partially observed systems through inexpensive sensors are facilitating real-time queue length estimations. 
%One of the main assumptions is knowing primary parameters such as flow rate and percent of connected vehicles (e.g., market penetration). 
The essential information needed for the estimation is primary parameters such as flow rate and percent of probe vehicles. However, both parameters are dynamic. Especially in real-time applications, like cycle-to-cycle or shorter-term queue lengths at signalized intersections, one would need to collect data for a few  cycles to estimate these parameters. 
%Then, they 
The parameters can then be updated and used in such applications. Assuming random arrivals, in (\cite{comert2016queue,comert2020cycle}), it is shown that 
%as low as 
at least 10 cycles of CV data would be needed to start using queue length estimators. 
%assuming random arrivals. 
%In this paper, we aimed to generally 
Our goal in this paper is to model queue lengths at intersections without assuming random arrivals or any primary parameters or estimating parameters. The problem is intuitive. The conditional probability of the location (order) of the last probe vehicle can be calculated by $P(L=l|M=m,N=n)=\binom{l-1}{m-1}/\binom{n}{m}$ given number of probe vehicles and the total number of vehicles in the queue. In this probability mass function, $L$ is the location of the last probe vehicle, $M$ is the number of probe vehicles in the queue, and $N$ is the total queue (\cite{comert2009}). We can see that this does not assume any arrival pattern or parameter and only depends on probe vehicle data.
%is for general case without assuming any arrival pattern or parameter only depending on CV data. 
%Certainly, it is not taking the advantage of time information $T$ arrival time of CVs with respect to signal timing. 
Certainly, it is not taking advantage of queue joining time $T$ of CVs with respect to signal timing. Also, we know that signal timing can be considered an integer (or increments of 5-10 seconds). There is also the physical constraint of the vehicle following, which can be 1-2 seconds $s$ even if vehicles arrive in a platoon.  
\begin{figure}[ht]
\centering
\includegraphics[width=0.6\linewidth]{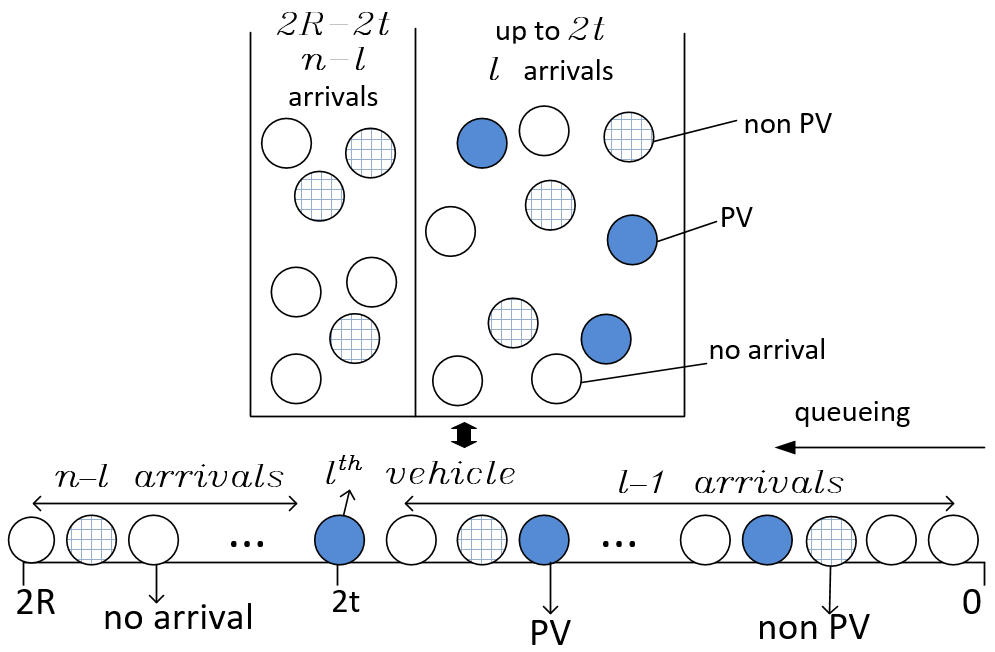}
\caption{Example queue with queue length, PV data, and arrival-no arrival spots}
\label{fig_1}       
\end{figure}

%In sum, the problem can be described on a single lane of an approach and equivalently express as urn modeling. 
The problem is described on a single lane of approach and can equivalently be expressed as an urn modeling (see Fig.~\ref{fig_1}). Consider the approach lane queue formed in $2R$ (i.e., $R$ is red phase) time intervals where in one time interval there can be at most one arrival. So, in this set up there can be at most one arrival per $0.5$ seconds. This can be thought of as the minimum possible time gap and can be updated in the formulations derived. In Fig.~\ref{fig_1}, we have $l$ arrivals in $2t$ time intervals among these $2t-1$ contain $m-1$ CVs, $2R-2t$ has $n-l$ arrivals. Now, the problem is a negative inference, meaning $n$ is changing as in Negative Binomial, so, we are interested in $P(N=n|L=l,M=m,T=t,R)$, i.e., probability of having $N=n$ arrivals within $R$ time interval given $L=l, T=t, M=m$. Calculating this probability, we obtain Eq.~(\ref{probability-1}) or equivalently Eq.~(\ref{probability-2}).
%But, this is simply calculating
\begin{equation}
\label{probability-1}
 \frac{\binom{2t}{l-m} \binom{2R-2t}{n-l}}{\binom{2R}{n-m}}
\end{equation}
% $\frac{\binom{2t}{l-m} \binom{2R-2t}{n-l}}{\binom{2R}{n-m}}$
\begin{equation}
\label{probability-2}
\frac{\binom{n-m}{l-m} \binom{2R-(n-m)}{2t-(l-m)}}{\binom{2R}{2t}}
\end{equation}
%$\frac{\binom{n-m}{l-m} \binom{2R-(n-m)}{2t-(l-m)}}{\binom{2R}{2t}}$ 
%and then calculating expected values (i.e., mean and variance) which will give us the queue length estimator ($E(n|l,t,m,R)$). However, we need to verify if 

We can then calculate expected values to get the mean ($E(n|l,t,m,R)$ or the queue length estimator) and the variance ($V(n|l,t,m,R)$) of the estimator. However, we first need to
\begin{enumerate}[i.]
\item verify if this is a valid probability mass function. 
\item find the normalizing denominator for a valid probability mass function.
\item simplify to forms that can be used as input-output models like $E(n|l,t,m,R)=l+(1-p)\lambda (R-t)$ in (\cite{comert2013simple}).
\item show if this 
%experiment simplifies 
approach leads to one of the known negative probability mass functions (e.g., Negative Hypergeometric). This could facilitate (iii).      
\end{enumerate}

\section{Probability Mass Function, Expected Value, and Variance}
\label{PEV}
We first  provide combinatorial arguments and derive a closed-form for the sum of the function in Eq. (\ref{probability-2}). \\
\begin{theorem}
\label{theorem1}
Let $\ell$, $t$, $R$, and $m$ be as defined in the preceding section. Then 
%$$\sum_{n=\ell}^{2R-2t+\ell}\frac{\binom{n-m}{\ell-m}\binom{2R+m-n}{2t+m-\ell}}%{\binom{2R}{2t}}
%=\frac{2R+1}{2t+1}$$ or equivalently
%$$\sum_{n=\ell}^{2R-2t+\ell}\binom{n-m}{\ell-m}\binom{2R+m-n}{2t+m-\ell}=\binom{2R+1}{2t+1}.$$ 
%\end{theorem}
%\bf Claim. \rm There is a closed-form of a normalizer for the function in Eq. %(\ref{probability-2}), and it can be shown that Eq.~(3) or equivalently Eq.~(4) %holds.
\begin{equation}
%\label{theorem1}
\sum_{n=\ell}^{2R-2t+\ell}\frac{\binom{n-m}{\ell-m}\binom{2R+m-n}{2t+m-\ell}}{\binom{2R}{2t}}
=\frac{2R+1}{2t+1}\\ 
\end{equation}
or equivalently
\begin{equation}
\sum_{n=\ell}^{2R-2t+\ell}\binom{n-m}{\ell-m}\binom{2R+m-n}{2t+m-\ell}=\binom{2R+1}{2t+1} 
\end{equation}

\end{theorem}
\bf Proof. \rm Observe $\binom{n-m}{\ell-m}=\binom{n-m}{n-\ell}$ and $\binom{2R+m-n}{2t+m-\ell}=\binom{2R+m-n}{2R-2t-n+\ell}$ and replace $n'=n-\ell$ so that Eq.~(4) becomes
\begin{eqnarray}
\label{step2}
\sum_{n'=0}^{2R-2t}\binom{n'+\ell-m}{n'}\binom{2R+m-n'-\ell}{2R-2t-n'}=\binom{2R+1}{2t+1}
\end{eqnarray}

Make another re-indexing $\ell'=\ell-m$ and hence Eq. (\ref{step2}) takes the form
\begin{eqnarray}
\label{step3}
\sum_{n'=0}^{2R-2t}\binom{n'+\ell'}{n'}\binom{2R-\ell'-n'}{2R-2t-n'}=\binom{2R+1}{2t+1}
\end{eqnarray}

The \it "negativization" \rm (reminiscent of the Euler's \it gamma reflection formula \rm $\Gamma(z)\Gamma(1-z)=\frac{\pi}{\sin(\pi z)}$) of binomial coefficients $\binom{-a+b}b=(-1)^b\binom{a-1}b$ allows to convert 
$\binom{\ell'+n'}{n'}=(-1)^{n'}\binom{-\ell'-1}{n'}$ and $\binom{2R-\ell'-n'}{2R-2t-n'}=\binom{2t-\ell'+2R-2t-n'}{2R-2t-n'}=(-1)^{2R-2t-n'}\binom{\ell'-2t-1}{2R-2t-n'}$.
Therefore,
\begin{eqnarray}
\label{step4}
\sum_{n'=0}^{2R-2t}\binom{n'+\ell'}{n'}\binom{2R-\ell'-n'}{2R-2t-n'}\nonumber \\ 
=\sum_{n'=0}^{2R-2t}\binom{-\ell'-1}{n'}\binom{\ell'-2t-1}{2R-2t-n'}
\end{eqnarray}

The well-known \it Vandermonde-Chu \rm identity states $\sum_{k=0}^y\binom{x}k\binom{z}{y-k}=\binom{x+z}y$. Applying this to Eq. (\ref{step4}) and engaging $\binom{-a+b}b=(-1)^b\binom{a-1}b$
(one more time) yields
$$\sum_{n'=0}^{2R-2t}\binom{-\ell'-1}{n'}\binom{\ell'-2t-1}{2R-2t-n'}=$$ $$\binom{-2t-2}{2R-2t}=(-1)^{2R-2t}\binom{2R+1}{2R-2t}=\binom{2R+1}{2t+1}$$
The proof is complete. $\square$

\bf Remark. \rm The identity just proved shows that Eq. (3) or Eq. (4) is independent of the parameters $m$ and $\ell$.
\newline

We can see that the identity proved in the above theorem enables us to revise Eq.~(\ref{probability-2}) and define a probability mass function. We can divide both sides of the identity by the expression on the right-hand side of the identity to get one on the right-hand side (i.e., the right-hand side of the identity is the normalizer of the probability distribution, which we explain below). Note that additional results from combinatorics and discussions are presented in the Appendix.  

In Negative Hypergeometric distribution (\cite{johnson2005univariate}), the probability of having $k$ successes up to the $r^{th}$ failure given sample size of $S$ and maximum possible queued vehicles $K$ is given by
\begin{equation}
\label{probneghyp}
p(k| r, K, S) = \frac{\binom{k+r-1}{k}\binom{S-r-k}{K-k}}{\binom{S}{K}}
\end{equation}
where $S$ is the sample size (time capacity for arrivals and non-arrivals), $K$ is the total number of successes (arrivals) in $S$, $r$ is the number of failures (non-arrivals), and $k$ is the number of successes (realizations of arrivals). The probabilities sum to $1$. For the Negative Hypergeometric distribution, the expected value $E(k|r,K,S)$ and the variance $V(k|r,K,S)$ are given by Eqs.~(\ref{expvalneghyp}) and (\ref{varneghyp}).
\begin{equation}
\label{expvalneghyp}
E(k|r,K,S) = \frac{rK}{S-K+1}
\end{equation}
\begin{equation}
\label{varneghyp}
V(k|r,K,S) =\frac{rK(S+1)(S-K-r+1)}{(S-K+1)^2(S-K+2)}
\end{equation}

In the probability mass function of the Negative Hypergeometric distribution (Eq.~(\ref{probneghyp})), let $S = 2R + 1$, $K = 2R - 2t$, 
$r = l - m + 1$, and $k = n - l$. Then the result proved in the theorem above gives the following probability mass function, which is a Negative Hypergeometric distribution since $\sum_{n=l}^{2R-2t+l} \binom{n - m}{l - m}\binom{2R + m - n}{2t + m - l}= \binom{2R + 1}{2t + 1}$. Notice that with these assignments arrivals and non-arrivals are fixed and probability of the total queue $N=n$ is calculated with known $l,m,t,R$. 
\begin{equation}
\label{probmassfn}
p(N=n|l,m,t,R) = \frac{\binom{n - m}{l - m}\binom{2R + m - n}{2t + m - l}}{\binom{2R + 1}{2t + 1}}
\end{equation}

From the formulas for the expected value and variance of the Negative Hypergeometric distribution,  we get the following formulas for the expected value (Eq.~(\ref{expvalneghyp})) and variance (Eq. (\ref{varneghyp})) of this probability distribution in Eq.~(\ref{probmassfn}).  Note that $L=l,M=m,T=t,R$ are basic information from CVs, not primary parameters (arrival or penetration rate of probe vehicle in the traffic stream). We also do not require steady-state behavior if this Probe vehicle information is available. The expected queue length and its variance are short-term ($R$ seconds or time interval) estimators. 

\begin{figure}[h!]
\captionsetup{aboveskip=3pt,belowskip=-5pt}
\captionsetup[subfigure]{aboveskip=2pt,belowskip=-1pt}
\centering
%\vspace{-10mm}
\begin{subfigure}{0.49\columnwidth}
\centering
  \includegraphics[width=0.99\columnwidth]{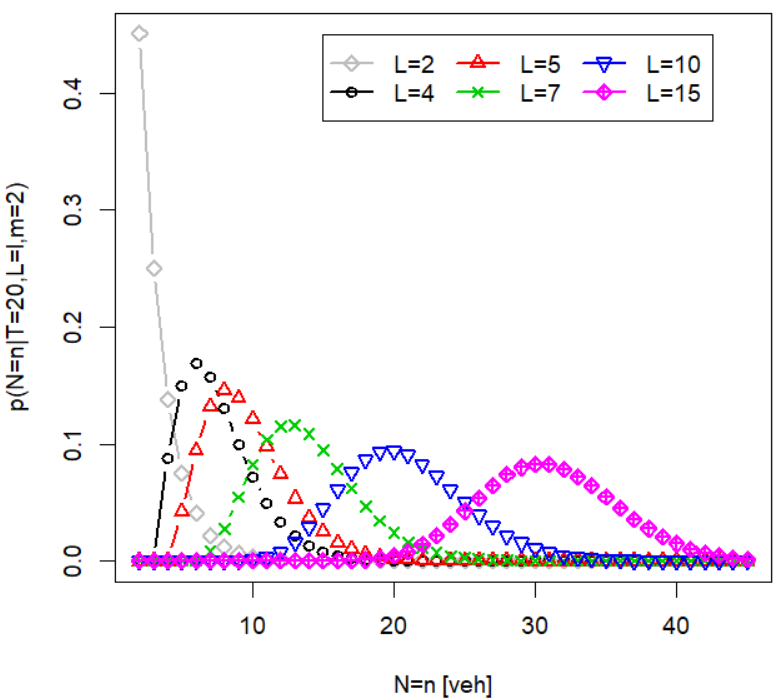}
\caption{$p(N|T=20,L,M=2,R=45)$ in Eq.~(\ref{probmassfn})}
  \label{fig_2a}
\end{subfigure} 
%\vspace{-3mm}
\captionsetup[subfigure]{aboveskip=2pt,belowskip=-1pt}
\begin{subfigure}{0.49\columnwidth}
 \centering
\includegraphics[width=0.99\columnwidth]{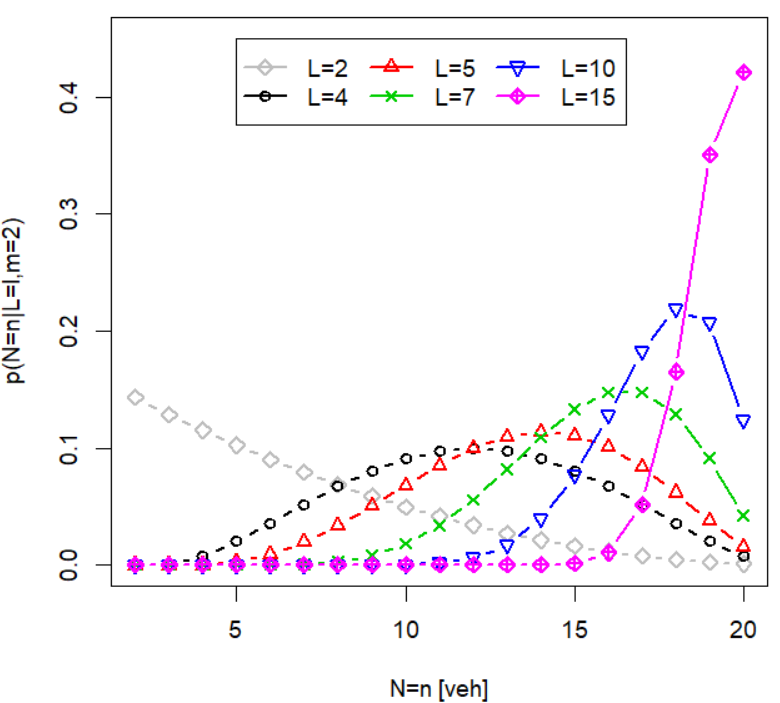}
\caption{$p(N|L,M=2,R=45)$ in Eq.~(\ref{probmassfn2})}
  \label{fig_2b}
\end{subfigure}
\caption{Example behavior of conditional probabilities}
\label{fig_2}
\end{figure}

The expected value $E(n|l,t,m,R)$ can be determined by 
%$E(n|l,t,m,R)=\sum_{n=l}^{2R-2t+l}{\frac{n(2R+1)}{(2t+1)}\frac{\binom{n-m}{l-m} \binom{2R+m-n}{2t+m-l}}{\binom{2R+1}{2t+1}}}=\sum_{n'=0}^{2R-2t}{\frac{n'(2R+1)}{(2t+1)}\frac{\binom{n'+l'}{n'} \binom{2R-l'-n'}{2R-2t-n'}}{\binom{2R+1}{2t+1}}}$ 
\begin{eqnarray*}
E(N|l,m,t,R)&=& \sum_{n=l}^{2R-2t+l}{\frac{n(2R+1)}{(2t+1)}\frac{\binom{n-m}{l-m} \binom{2R+m-n}{2t+m-l}}{\binom{2R+1}{2t+1}}}\\
&=& \sum_{n'=0}^{2R-2t}{\frac{n'(2R+1)}{(2t+1)}\frac{\binom{n'+l'}{n'} \binom{2R-l'-n'}{2R-2t-n'}}{\binom{2R+1}{2t+1}}} 
\end{eqnarray*}
where $n'=n-l$, $l'=l-m$, and $\frac{(2R+1)}{(2t+1)}$ is the normalizer. 
%\begin{eqnarray}
%\label{eqn_nhg1}
%E(N_1|l,m,t,R) &= & l+\frac{(l-m+1)(2R-2t)}{2t+2} \nonumber\\
%&=& l+\frac{(l-m+1)(R-t)}{t+1} \nonumber\\
%V(N_1|l,m,t,R) 
%&=& \frac{(l-m+1)(2R+2)(2R-2t)}{(2t+2)(2t+3)}[1-\frac{l-m+1}{2t+2}]
%\end{eqnarray} 

By Eqs. (\ref{expvalneghyp}) and (\ref{varneghyp}), simplified expected value or the queue length estimation 1 and the variance can be obtained as in Eqs.~(\ref{eqn_nhg1}) and (\ref{eqn_vnhg1}).
\begin{eqnarray}
\label{eqn_nhg1}
E(N_1=n_1|l,m,t,R) &= & l+\frac{(l-m+1)(2R-2t)}{2t+2}\nonumber \\ 
&=& l+\frac{(l-m+1)(R-t)}{t+1}
\end{eqnarray}
\begin{eqnarray}
\label{eqn_vnhg1}
V(N_1=n_1|l,m,t,R)=&\frac{(l-m+1)(2R+2)(2R-2t)}{(2t+2)(2t+3)}
[1-\frac{l-m+1}{2t+2}]
\end{eqnarray} 
%It is not intuitive to see expected value, however, we are simply calculating

Alternatively, from Eq.~(\ref{probmassfn2}), we can get the following 
%second 
equivalent estimator without CV time  ($T$) information (Eq.~(\ref{eqn_nhg2})) and its variance in (Eq.~(\ref{eqn_vnhg2})). 
%$C$ is capacity or maximum possible arrivals, $J=C-2l$, $r=l-m+1$, $K=C-l$, and $k=n-l$. Note that, with time discretization we are able to infer $t$ from $l$.
\begin{equation}
\label{probmassfn2}
p(N=n|l, m, C) = \frac{\binom{C-n+m}{C-n} \binom{n-m}{n-l}}{\binom{C}{l}}
\end{equation}
where $C$ is capacity or maximum possible arrivals, $J=C-2l$, $r=l-m+1$, $K=C-l$, and $k=n-l$. Note that, with time discretization, we can infer $t$ from $l$.
%It is not intuitive to see expected value, however, we are simply calculating $E(n|l,m,R)=\sum_{n=l}^{C+l}{\frac{n(C+1)}{(l+1)}\frac{\binom{C-n+m}{C-n} \binom{n-m}{n-l}}{\binom{C}{l}}}$=$\sum_{n'=0}^{C}{\frac{n'(C+1)}{(l+1)}\frac{\binom{n'+l'}{n'} \binom{C-l'+n'}{C-l'-n'}}{\binom{C}{n'}}}$ with $n'=n-l$, $l'=l-m$, and normalizer $\frac{(C+1)}{(l+1)}$.  
%It is not intuitive to see expected value, however, we are simply 
%calculating 
The expected value $E(n|l,m,R)$ is given by
\begin{eqnarray*}
E(N=n|l,m,R) &=& \sum_{n=l}^{C+l}{\frac{n(l+1)}{(C+1)}\frac{\binom{C-n+m}{C-n} \binom{n-m}{n-l}}{\binom{C}{l}}} \\
&=& \sum_{n'=0}^{C}{\frac{n'(l+1)}{(C+1)}\frac{\binom{n'+l'}{n'} \binom{C-l'+n'}{C-l'-n'}}{\binom{C}{n'}}} 
\end{eqnarray*}
where $n'=n-l$, $l'=l-m$, and $\frac{(l+1)}{(C+1)}$ is the normalizer for valid probability mass function.
\begin{equation}
\label{eqn_nhg2}
E(N_2=n_2|l,m,C)=l+\frac{(l-m+1)(C-l)}{l+2}
\end{equation}
\begin{equation}
\label{eqn_vnhg2}
V(N_2=n_2|l,m,C) =\frac{(l-m+1)(C+2)(C-l)}{(l+2)(l+3)}[1-\frac{l-m+1}{l+2}]
\end{equation}

\begin{figure}[h!]
\captionsetup{aboveskip=3pt,belowskip=-5pt}
\captionsetup[subfigure]{aboveskip=2pt,belowskip=-1pt}
\centering
%\vspace{-10mm}
\begin{subfigure}{0.49\columnwidth}
\centering
  \includegraphics[width=0.99\columnwidth]{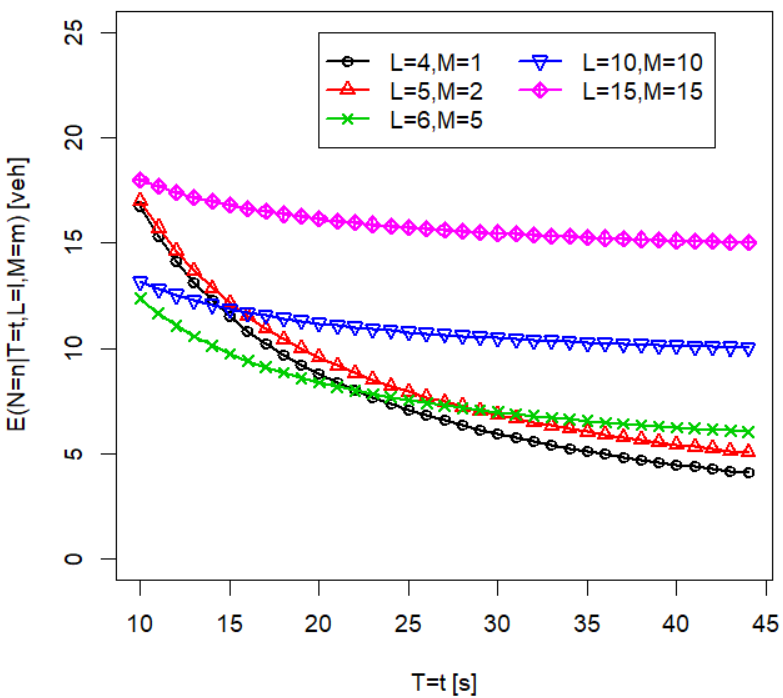}
\caption{$E(N|T,L,M=2,R=45)$ in Eq.~(\ref{eqn_nhg1})}
  \label{fig_3a}
\end{subfigure} 
%\vspace{-3mm}
\captionsetup[subfigure]{aboveskip=2pt,belowskip=-1pt}
\begin{subfigure}{0.49\columnwidth}
 \centering
\includegraphics[width=0.99\columnwidth]{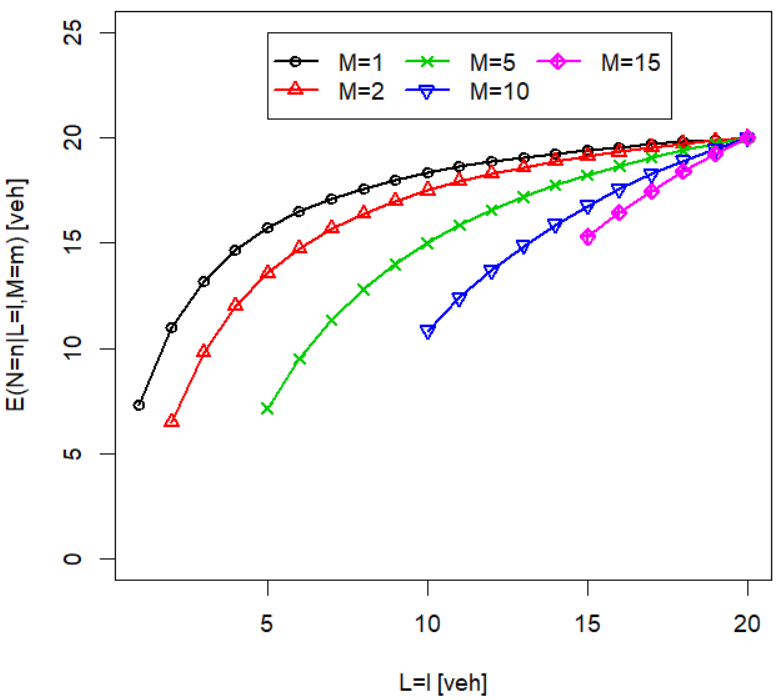}
\caption{$E(N|L,M,R=45)$ in Eq.~(\ref{eqn_nhg2})}
  \label{fig_3b}
\end{subfigure}
\caption{Example behavior of conditional expectations}
\label{fig_3}
\end{figure}

\begin{figure}[h!]
\captionsetup{aboveskip=3pt,belowskip=-5pt}
\captionsetup[subfigure]{aboveskip=2pt,belowskip=-1pt}
\centering
%\vspace{-10mm}
\begin{subfigure}{0.49\columnwidth}
\centering
  \includegraphics[width=0.99\columnwidth]{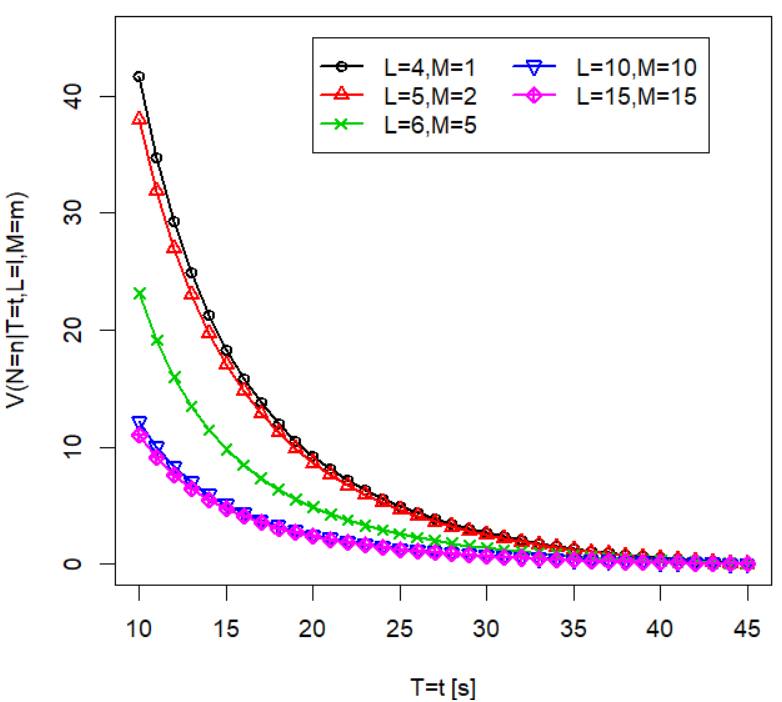}
\caption{$V(N|T,L,M=2,R=45)$ in Eq.~(\ref{eqn_nhg1})}
  \label{fig_4a}
\end{subfigure} 
%\vspace{-3mm}
\captionsetup[subfigure]{aboveskip=2pt,belowskip=-1pt}
\begin{subfigure}{0.49\columnwidth}
 \centering
\includegraphics[width=0.99\columnwidth]{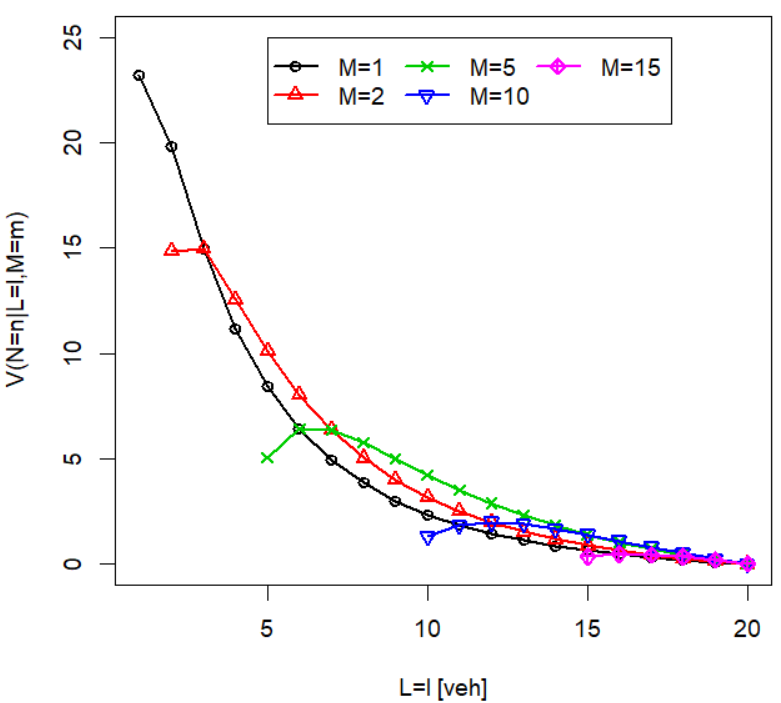}
\caption{$V(N|L,M,R=45)$ in Eq.~(\ref{eqn_vnhg2})}
  \label{fig_4b}
\end{subfigure}
\caption{Example behavior of conditional variances}
\label{fig_4}
\end{figure}

One of the advantages of the derived estimators in Eqs.~(\ref{eqn_nhg1}) and (\ref{eqn_nhg2}) is that the denominators are nonzero since $L\geq0$. This enables us to estimate queues even if there is no probe vehicle in the queue. The behavior of conditional probabilities, expected values, and variances are shown in Figs.~\ref{fig_2}-\ref{fig_4}. We can see in Fig.~\ref{fig_2} that the likelihoods are right to the $N=l$ values. In Fig.~\ref{fig_3}, as queue time joining of the last probe vehicle increases, then the expected queue length gets closer to $L=l$ for both models. Similarly, in Fig.~\ref{fig_4}, the variance of the estimated queue length reduces as $l$ and $t$ increase. Having time information also shows smoother behavior compared to having only location information.     


\section{Evaluation with Field Queue Length Data}
\label{field}
%In order to 
To show the effectiveness of the estimators developed, 
%we compared them using 
we used 2014 ITS World Congress Connected Vehicle Demonstration Data (\cite{dataset}). Authors' previous works used this field data for evaluating range sensor inclusion and filtering for queue length estimation (\cite{comert2020cycle,comert2021queue}). Results for this study are new. For completeness, assumptions and set up are reported again. The dataset contains manually collected queue lengths at the intersection of Larned and Shelby streets in Detroit, Michigan, between September $8$ and $10$, $2014$. The number of observations per day are $98$, $254$, and $135$, respectively. During data collection, probe vehicles were identified with the blue $X$s. Each row of data includes hour, minute, second of an observation, the maximum queue lengths, and the number of probe vehicles in these queues (i.e., $M$ in the formulations above) from left, center, and right lanes of Larned street approach.
 
The dataset provides $M=m$ and $C$ cycle time values but not the information of $L$ and $T$ 
%information 
from PVs. Hence, we generated random variates of this information from Uniform ($L=l$ location $l\sim U(m,n)$) and Gamma ($T=t$ queue joining time $t\sim\mathcal{G}a(l,\frac{C}{2n})$) distributions for all lanes independently and repeated for $1000$ random seeds. Note that, integer values are used for $L,M$, and $T$. Overall average of estimation errors are reported to compare models. In addition, followings are assumed related to the traffic signal and the dataset:
\begin{enumerate}
\item Back of queue observations are obtained at the end of red phases (vary cycle-by-cycle). The time between two observations is assumed to be the cycle length ($C$) and red phases are assumed to be half ($R=C/2$). 
\item There is no steady growth of queue and many zero queue values. Thus, the overflow queues are omitted . The data was collected during low to medium $\rho$ (i.e., volume-to-capacity ratio$=0.50$). Regardless, $\rho$s are also calculated using estimated arrival rates. 
\item The capacity of the approach was approximated by the observed overall maximum queue value of $10$ vehicles within $70$ seconds ($10\times3600/35$=$1029$ vehicles per hour or $0.286$ vehicles per second ($vps$) saturation flow rate). These values are used essentially in the Highway Capacity Manual (HCM) from a manual and back of queue calculations. Note that the values may not be reflecting actual capacity and phase splits; 
however, we compare and report against true queue lengths. This would provide insights into the accuracy of our approach.
\end{enumerate}

Compared HCM delay (i.e., $Delay$) and back of queue (i.e., $Q {back}$) models are given in Eqs. (\ref{eqn_hcm}) and (\ref{eqn_qb}). These models are approximations for given time intervals (e.g., $15$ minutes) and fully observed traffic.
\begin{eqnarray}
\label{eqn_hcm}
d_1=\frac{C}{2}[\frac{(1-G/C)^2}{1-[min(1,\hat{X})G/C]}]\nonumber\\
d_2=900T[(\hat{X}-1)+\sqrt{(\hat{X}-1)^2+\frac{8kI\hat{X}}{cT}}]
\end{eqnarray}
where $d$=$d_1\times PF+d_2+d_3$ is control delay seconds per vehicle, $d_2$ is uniform delay, $PF$ is progression factor due to arrival types, $d_2$ is random delay component, and $d_3$ is delay due to initial queue. In this study, only $d_1+d_2$ are considered with $d_3=0$ since no overflow queue is assumed. $PF=1.0$ is used for random arrivals. Volume-to-capacity is $\hat{X}=\hat{\rho}=\frac{\hat{\lambda}}{0.286}$. Green time $G$ is in seconds $s$, $C$ is cycle time in $s$. $T$ is the analysis period in hours where in cycle-to-cycle estimations $T_i=C_i/3600$ is assumed where $i$ denotes cycle number. $k$ is incremental delay factor, and $0.5$ is assumed for fixed time like movement. $I=1$ upstream filtering is assumed for no interaction with nearby intersections, and capacity is $c=1029$ $vph$. Note that in our calculations, uniform delay is the main component updated by changing $G$ and $C$ values. Queue lengths are approximated by Little's formula $d\lambda$ where $d$ and  $\lambda$ are both calculated at each cycle using $M$ number of probe vehicles in the queue. This method is based on HCM 2000 (\cite{prassas2020highway, ni2020signalized}).

Another estimation approach adopted from (\cite{kyte2014operation}) that is used to calculate cycle-to-cycle back of queues (see Eq.~(\ref{eqn_qb})). 
\begin{eqnarray}
\label{eqn_qb}
Q_{back}=\hat{v}(R+g_s)
\end{eqnarray}     
where, $Q_{back}$ is back of the queue in vehicles, $v=\lambda$ is arrival rate in vehicles per second ($vps$), $R$ is the red duration in seconds $s$, and $g_s$ is queue service time that is calculated $\hat{v}R/(x-\hat{v})$ with $x$ is the saturation flow rate (i.e., assumed to be $0.286$ vps). All the values $R$, $g_s$, and $\hat{v}$ except $x$ are changing cycle-to-cycle.

\begin{figure}[h!]
\captionsetup{aboveskip=3pt,belowskip=-5pt}
\captionsetup[subfigure]{aboveskip=2pt,belowskip=-1pt}
\centering
%\vspace{-10mm}
\begin{subfigure}{0.49\columnwidth}
\centering
  \includegraphics[width=0.99\columnwidth]{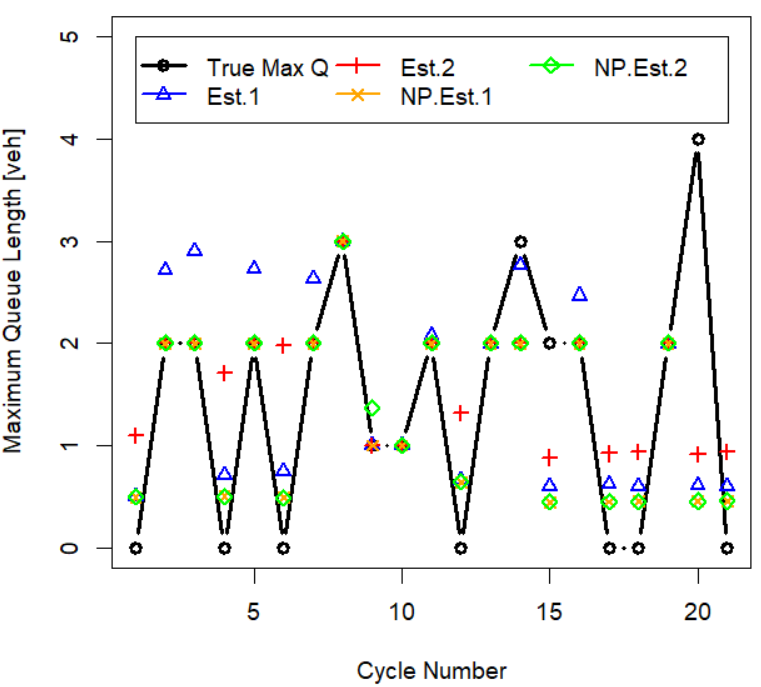}
\caption{Estimation on Sep 08 on center-lane}
  \label{fig_est}
\end{subfigure} 
%\vspace{-3mm}
\captionsetup[subfigure]{aboveskip=2pt,belowskip=-1pt}
\begin{subfigure}{0.49\columnwidth}
 \centering
\includegraphics[width=0.99\columnwidth]{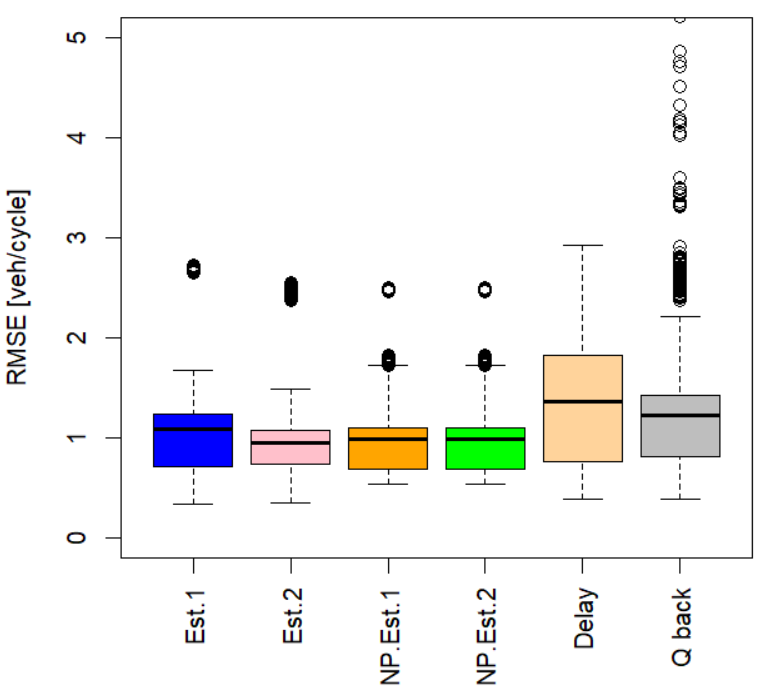}
\caption{Box plots for all estimation errors}
  \label{fig_box}
\end{subfigure}
\caption{Performance of the proposed estimators $NP.Est.1$ and $NP.Est.2$}
\label{fig_5}
\end{figure}

Alternative estimators from \cite{comert2016queue} are denoted by Est.1 and Est.2 in Eqs.~(\ref{est.1}) and (\ref{est.2}), respectively. These queue length estimators are in the form of $E(n|l,m,t,R)=l+(1-\hat{p})\hat{\lambda}(R-t)$ with two different primary parameter estimator combinations: $\{\hat{\lambda}_1=\frac{l}{R}, \hat{p}_1=\frac{m}{l}\}$ and $\{\hat{\lambda}_2=\frac{(l-m)}{t}+\frac{m}{R}, \hat{p}_2=\frac{mt}{mt+(l-m)R}\}$.
\begin{equation}
\label{est.1}
E(N_1|l,m,t,R)=I(m>0)[l+(l-m)(1-\frac{t}{R})]+I(m=0)[(1-\frac{\bar{m}}{\bar{l}})(\bar{l}+(\bar{l}-\bar{m})(1-\frac{\bar{t}}{R}))]    
\end{equation}
\begin{eqnarray}
\label{est.2}
E(N_2|l,m,t,R)=I(m>0)[m+\frac{R(l-m)}{t}]+I(m=0)[(1-\frac{\bar{m}\bar{t}}{\bar{m}\bar{t}+(\bar{l}-\bar{m})R})(\bar{m}+\frac{R(\bar{l}-\bar{m})}{\bar{t}})] \nonumber \\
\centering
=I(m>0)[m+\frac{(l-m)R}{t}]+I(m=0)[\bar{m}+\frac{(\bar{l}-\bar{m})R}{\bar{t}}]
\end{eqnarray}
where $I(.)$ is the indicator function. When there is no probe vehicle in the queue (i.e., $I(m=0)$), we use the average of previous probe vehicles' information as we need to estimate arrival rate ($\lambda$) and probe percentage ($p$). Notation $M_{1:i}$ represents values from cycle $1$ to $i$ and $\bar{m}_{1:i}=\sum_{j=1}^{i}{\frac{m_j}{i}}$, $\bar{l}_{1:i}=\sum_{j=1}^{i}{\frac{l_j}{i}}$, and $\bar{t}_{1:i}=\sum_{j=1}^{i}{\frac{t_j}{i}}$. Average error values are given in Table~\ref{tab_data} for $T\sim\mathcal{G}a(l,\frac{C}{2n})$. Fig.~\ref{fig_box} is given to demonstrate if assumed interarrivals are impacting the accuracy of the estimators.

\begin{table}[h!]
	\centering	
	\caption{\textcolor{black}{Estimation results with RMSE errors in [vehs/cycle] with $T\sim\mathcal{G}a(l,$C$/(2n))$}}
	\label{tab_data}       % Give a unique label
	%\resizebox{\textwidth}{!}{%
	\scalebox{0.85}{
		\begin{tabular}{c c c c c c c cc}
			\hline\noalign{\smallskip}
			& Lane&  Avg. $p$ & Est.1 & Est.2 & NP.Est.1 & NP.Est.2 & Delay & Q back \\
			 \hline\noalign{\smallskip}
			
			\multirow{3}{*}{Sep08}
			& L & 13\%	&1.09	&1.01	& 1.01	& 1.01	& 1.37	& 1.25	\\ %\hline
			& C & 21\%	&0.72	&0.78	&0.69	&0.70	&1.33	&1.26	\\ %\hline
			& R & 7\%	&0.56	&0.55	&0.60	&0.60	&0.66	&0.64	\\ %\hline
			\hline
			\multirow{3}{*}{Sep09} 
			& L & 10\%	&1.22	&1.05	&0.99	&0.99	&1.38	&1.13	\\ %\hline
			& C & 26\%	&1.14	&0.96	&1.09	&1.09	&1.81	&1.38	\\ %\hline
			& R & 2\%	&0.34	&0.35	&0.54	&0.54	&0.39	&0.41	\\ %\hline
			\hline
			\multirow{3}{*}{Sep10} 
			& L & 7\%	&2.68	&2.43	&2.48	&2.48	&2.81	&2.52	\\ %\hline
			& C & 18\%	&1.48	&1.32	&1.73	&1.73	&2.28	&1.63	\\ %\hline
			& R & 1\%	&0.84	&0.77	&0.77	&0.77	&1.06	&1.26	\\ %\hline			
			\noalign{\smallskip}\hline\noalign{\smallskip}
		\end{tabular}		
	}
\end{table}  
In Table~\ref{tab_data}, a summary of average queue length ($QL$) estimation errors in the root mean square is provided 
%in root mean squared errors
(RMSE=$\sqrt{\sum_{i=1}^{n}{\frac{(QL_i-\hat{QL}_i)^2}{n}}}$). Average $p$ values are calculated from $\sum_{i=1}^{n}{\frac{m}{nQL_i}}$ for each lane. Since true maximum queues are not known, $p$ and $\lambda$ are estimated. HCM's control delay-based model and back of queue are denoted by $HCM_d$ and $Q_{back}$, respectively. The accuracy of the estimators is reported when probe vehicles are present in the queue ($p$=$\{10\%, 13\%, 18\%, 21\%$, $26\%\}$). 

Example performances with $21\%$ penetration rates is given in Fig.~\ref{fig_est}. When there are probe vehicles in the queue, we can see that the proposed methods can follow the true maximum queue lengths closely. In Fig.~\ref{fig_box}, boxplots for overall errors are given. We can see that the model with new estimators provides slightly lower errors. However, errors are lower than delay-based $HCM_d$ and $Q_{back}$ methods. Our methods can estimate more accurately compared to $Q_{back}$.
%In Fig.~\ref{fig_est}, mixed results are obtained up to $p=10\%$ with some negative improvements. Estimators with range sensor clearly perform better as $p$ increases. From Fig.~\ref{fig_errsp}-b, a trend of higher improvement of $Est2$ over $Q~back$ method with respect to $p$ can be seen.

\section{Conclusions}
\label{sctconc}
In this study, we derived two new nonparametric queue length estimation models for traffic signal-induced queues. The estimators only depend on signal phasing and timing information. The derivations involved fundamental experiment setup, and 
the resulting 
estimators were simple algebraic expressions. We did not assume independent arrivals at the intersection. The only assumption we made was discrete time intervals which are 
%quite relevant 
reasonable as signal timing involves whole seconds; in fact, multiples of five seconds. 

For independent approach lanes at traffic intersections, it is shown that conditional queue lengths given probe vehicle location, count, time, and analysis interval can be represented by a Negative Hypergeometric distribution. 
The performance of the estimators derived
%Derived estimators' performance 
was compared with
%against 
parametric and simple highway capacity manual methods that use field test data involving probe vehicles. The results obtained from the comparisons show that the nonparametric models presented in this paper match the accuracy of parametric models. 
The methods developed do not assume random arrivals of vehicles at the intersection or any primary parameters or involve parameter estimations.

%In the study, we did not aim to improve the performance of the existing methods. 
In this study, we developed methods to estimate queue length at intersection approaches from probe vehicles. These probe vehicles could be traditional probe vehicles or connected vehicles that generate basic safety messages. Future research could study the models presented in this paper to a more complex intersection and a series of adjacent intersections with large traffic demand volume at these intersections.
%estimators more general nonparametric queue length estimators from probes. Therefore, limitations of this study include that (1) applications could be applied to complex series of intersections  and (2) a larger dataset could be used to show the effectiveness of the proposed methods.
%estimating parameters. 

%Based on the data used, nonparametric models can match the accuracy of parametric models.

%\begin{enumerate}
%\item 
%\end{enumerate}
% \vspace{-1.5pt}
%Results from numerical analysis also revealed the accuracy of the nonparametric %estimators are able to match parametric ones.
\vspace{-10pt}
\section*{Acknowledgments}
This study is partially supported by the Center for Connected Multimodal Mobility ($C^{2}M^{2}$) (USDOT Tier 1 University Transportation Center) headquartered at Clemson University, Clemson, SC. Any opinions, findings, and conclusions or recommendations expressed in this paper are those of the authors and do not necessarily reflect the views of $C^{2}M^{2}$ and the official policy or position of the USDOT/OST-R, or any State or other entity, and the U.S. Government assumes no liability for the contents or use thereof. It is also partially supported by U.S. Department of Homeland Security SRT Follow-On grant, Department of Energy-National Nuclear Security Administration (NNSA) PuMP, MSIPP  IAM-EMPOWEREd, MSIPP, Department of Education MSEIP programs, NASA ULI (University of South Carolina-Lead), and NSF Grant Nos. 1719501, 1954532, and 2131080.
 \vspace{-5pt}
 
\section*{Appendix}
\label{append}
\noindent
The results in Eq. (\ref{theorem1}) can be extended to the short sum runs from $n=\ell$ through $n=2R-2t$.
\smallskip
\begin{theorem}
\label{theorem2}
Let $\ell$, $m$, $n$, $R$, and $t$ be as defined in Theorem \ref{theorem1}. Then  \it there is a recurrence formula for
$$\sum_{n=\ell}^{2R-2t}\binom{n-m}{\ell-m}\binom{2R+m-n}{2t+m-\ell}. \quad (5)$$ \rm
\end{theorem}
\bf Proof. \rm Denote the sum by $f(\ell)$ and the summand by $F(\ell,n)$ (after suppressing the remaining variables). Introduce the function
$G(\ell,n)=-\binom{n-m}{\ell+1-m}\binom{2R+m-n+1}{2t+m-\ell}$. Then, it is routine to verify that
$$F(\ell+1,n)-F(\ell,n)=G(\ell,n+1)-G(\ell,n). \quad (6)$$
Sum both sides of (6) for $n=\ell+1$ to $n=2R-2t$ (and telescoping on the right-hand side) to obtain
$$f(\ell+1)-f(\ell)+F(\ell,\ell)=G(\ell,2R-2t+1)-G(\ell,\ell+1).$$
Based on $F(\ell,\ell)=\binom{2R+m-\ell}{2t+m-\ell}, G(\ell,2R-2t+1)=-\binom{2R-2t-m+1}{\ell-m+1}\binom{2t+m}{\ell}$ and $G(\ell,\ell+1)=-\binom{2R+m-\ell}{2t+m-\ell}$, we infer the
\bf recursive relation \rm
$$f(\ell+1)-f(\ell)=-\binom{2R-2t-m+1}{\ell-m+1}\binom{2t+m}{\ell}. \qquad \square$$
\bf Corollary. \it From Theorem \ref{theorem2}, we get the following identity
$$\sum_{\ell=0}^{2R-2t}\binom{2R-2t-m+1}{\ell-m+1}\binom{2t+m}{\ell}=\binom{2R+1}{2t+1}. \quad (7)$$ \rm
\bf Proof. \rm This follows from the recurrence relation proved in Theorem \ref{theorem2} and the identity proved in Theorem \ref{theorem1}. $\square$
\bigskip

\noindent
\begin{theorem}
\label{theorem3}
%\bf Remark. \rm Actually, 
The identity in (7) can be re-indexed and formulated as follows: 
%and slightly generalized to formulate as
$$\sum_{m=\ell}^{R-t+\ell } {m\choose \ell}{R-m\choose t-\ell}={R+1\choose t+1}.$$
\end{theorem}
\bf Proof. \rm We offer a combinatorial argument.
Given natural numbers $\ell\le t\le R$, and $m$, we may consider the class of those $(t+1)$-subsets $\{x_0<x_1<\dots<x_t\}$ of $\{0,1,\dots,R\}$ such that $x_\ell=m$: these are exactly ${m\choose \ell}{R-m\choose t-\ell}$ (indeed the $\ell$ elements $x_0,\dots, x_{\ell-1}$ can be chosen freely into $\{0,\dots, m-1\}$, and so can the $t-\ell$ elements $x_{\ell+1},\dots,x_t$ into $\{m+1,\dots,R\}$. These classes, for $ \ell\le m\le R-t+\ell $ form a partition of all $(t+1)$-subsets of $[R+1]$, whence the sum of their cardinality is independent of $\ell$ and the identity.
$\square$

\bigskip
\noindent
\bf Remark. \rm The discrepancy in having a closed form and no closed form can be understood as follows: we know that $\sum_{k=0}^n\binom{n}k=2^n$, however there is no "nice evaluation" for $\sum_{k=0}^m\binom{n}k$ unless $m=n$. The bottom line is the former is summed over the full compact support of $\binom{n}k$ (in the sense, $\binom{n}k=0$ if $k<0$ or $k>n$. A similar analogy can be drawn with having the closed form $\int_{\Bbb{R}}e^{-x^2}dx=\sqrt{\pi}$ but nothing similar is available if the limit are altered to be any smaller subset than the full range $\Bbb{R}$, except for $[0,\infty)$.

\bibliography{detection_trb2}

\begin{thebibliography}{34}
\expandafter\ifx\csname natexlab\endcsname\relax\def\natexlab#1{#1}\fi
\providecommand{\url}[1]{\texttt{#1}}
\providecommand{\href}[2]{#2}
\providecommand{\path}[1]{#1}
\providecommand{\DOIprefix}{doi:}
\providecommand{\ArXivprefix}{arXiv:}
\providecommand{\URLprefix}{URL: }
\providecommand{\Pubmedprefix}{pmid:}
\providecommand{\doi}[1]{\href{http://dx.doi.org/#1}{\path{#1}}}
\providecommand{\Pubmed}[1]{\href{pmid:#1}{\path{#1}}}
\providecommand{\bibinfo}[2]{#2}
\ifx\xfnm\relax \def\xfnm[#1]{\unskip,\space#1}\fi
%Type = Article
\bibitem[{Bae et~al.(2019)Bae, Liu, Han and Bozdogan}]{bae2019spatio}
\bibinfo{author}{Bae, B.}, \bibinfo{author}{Liu, Y.}, \bibinfo{author}{Han,
  L.D.}, \bibinfo{author}{Bozdogan, H.}, \bibinfo{year}{2019}.
\newblock \bibinfo{title}{Spatio-temporal traffic queue detection for
  uninterrupted flows}.
\newblock \bibinfo{journal}{Transportation Research Part B: Methodological}
  \bibinfo{volume}{129}, \bibinfo{pages}{20--34}.
%Type = Article
\bibitem[{Comert(2013a)}]{Comert2013}
\bibinfo{author}{Comert, G.}, \bibinfo{year}{2013}a.
\newblock \bibinfo{title}{Effect of stop line detection in queue length
  estimation at traffic signals from probe vehicles data}.
\newblock \bibinfo{journal}{European Journal of Operational Research}
  \bibinfo{volume}{226}, \bibinfo{pages}{67--76}.
%Type = Article
\bibitem[{Comert(2013b)}]{comert2013simple}
\bibinfo{author}{Comert, G.}, \bibinfo{year}{2013}b.
\newblock \bibinfo{title}{Simple analytical models for estimating the queue
  lengths from probe vehicles at traffic signals}.
\newblock \bibinfo{journal}{Transportation Research Part B: Methodological}
  \bibinfo{volume}{55}, \bibinfo{pages}{59--74}.
%Type = Article
\bibitem[{Comert(2016)}]{comert2016queue}
\bibinfo{author}{Comert, G.}, \bibinfo{year}{2016}.
\newblock \bibinfo{title}{Queue length estimation from probe vehicles at
  isolated intersections: Estimators for primary parameters}.
\newblock \bibinfo{journal}{European Journal of Operational Research}
  \bibinfo{volume}{252}, \bibinfo{pages}{502--521}.
%Type = Article
\bibitem[{Comert and Begashaw(2021)}]{comert2020cycle}
\bibinfo{author}{Comert, G.}, \bibinfo{author}{Begashaw, N.},
  \bibinfo{year}{2021}.
\newblock \bibinfo{title}{Cycle-to-cycle queue length estimation from connected
  vehicles with filtering on primary parameters}.
\newblock \bibinfo{journal}{International Journal of Transportation Science and
  Technology} \URLprefix
  \url{https://www.sciencedirect.com/science/article/pii/S2046043021000319},
  \DOIprefix\doi{https://doi.org/10.1016/j.ijtst.2021.04.009}.
%Type = Article
\bibitem[{Comert and Cetin(2009)}]{Comert2009}
\bibinfo{author}{Comert, G.}, \bibinfo{author}{Cetin, M.},
  \bibinfo{year}{2009}.
\newblock \bibinfo{title}{Queue length prediction from probe vehicle location
  and the impacts of sample size}.
\newblock \bibinfo{journal}{European Journal of Operational Research}
  \bibinfo{volume}{197}, \bibinfo{pages}{196--202}.
%Type = Article
\bibitem[{Comert and Cetin(2021)}]{comert2021queue}
\bibinfo{author}{Comert, G.}, \bibinfo{author}{Cetin, M.},
  \bibinfo{year}{2021}.
\newblock \bibinfo{title}{Queue length estimation from connected vehicles with
  range measurement sensors at traffic signals}.
\newblock \bibinfo{journal}{Applied Mathematical Modelling}
  \bibinfo{volume}{99}, \bibinfo{pages}{418--434}.
%Type = Misc
\bibitem[{Dataset(2014)}]{dataset}
\bibinfo{author}{Dataset, C.}, \bibinfo{year}{2014}.
\newblock \bibinfo{title}{{ITS} {W}orld {C}ongress {C}onnected {V}ehicle {T}est
  {B}ed {D}emonstration {V}ehicle {S}ituation {D}ata}.
\newblock \bibinfo{howpublished}{South East Michigan Test Bed Contractor Team,
  Noblis’ Queue Length Algorithm Development Team, and Data Capture and
  Management Data Sets Contractor Team, provided by ITS DataHub through
  Data.transportation.gov}.
\newblock \URLprefix \url{Accessed 2021-01-25 from
  http://doi.org/10.21949/1504496}.
%Type = Article
\bibitem[{Duret and Yuan(2017)}]{duret2017traffic}
\bibinfo{author}{Duret, A.}, \bibinfo{author}{Yuan, Y.}, \bibinfo{year}{2017}.
\newblock \bibinfo{title}{Traffic state estimation based on eulerian and
  lagrangian observations in a mesoscopic modeling framework}.
\newblock \bibinfo{journal}{Transportation research part B: methodological}
  \bibinfo{volume}{101}, \bibinfo{pages}{51--71}.
%Type = Article
\bibitem[{Florin and Olariu(2020)}]{florin2020towards}
\bibinfo{author}{Florin, R.}, \bibinfo{author}{Olariu, S.},
  \bibinfo{year}{2020}.
\newblock \bibinfo{title}{Towards real-time density estimation using
  vehicle-to-vehicle communications}.
\newblock \bibinfo{journal}{Transportation research part B: methodological}
  \bibinfo{volume}{138}, \bibinfo{pages}{435--456}.
%Type = Article
\bibitem[{Gao et~al.(2019)Gao, Han, Dong, Xiong and Du}]{gao2019connected}
\bibinfo{author}{Gao, K.}, \bibinfo{author}{Han, F.}, \bibinfo{author}{Dong,
  P.}, \bibinfo{author}{Xiong, N.}, \bibinfo{author}{Du, R.},
  \bibinfo{year}{2019}.
\newblock \bibinfo{title}{Connected vehicle as a mobile sensor for real time
  queue length at signalized intersections}.
\newblock \bibinfo{journal}{Sensors} \bibinfo{volume}{19},
  \bibinfo{pages}{2059}.
%Type = Article
\bibitem[{Goldenshluger(2016)}]{goldenshluger2016nonparametric}
\bibinfo{author}{Goldenshluger, A.}, \bibinfo{year}{2016}.
\newblock \bibinfo{title}{Nonparametric estimation of the service time
  distribution in the m/g/∞ queue}.
\newblock \bibinfo{journal}{Advances in Applied Probability}
  \bibinfo{volume}{48}, \bibinfo{pages}{1117--1138}.
%Type = Article
\bibitem[{Goldenshluger and Koops(2019)}]{goldenshluger2019nonparametric}
\bibinfo{author}{Goldenshluger, A.}, \bibinfo{author}{Koops, D.T.},
  \bibinfo{year}{2019}.
\newblock \bibinfo{title}{Nonparametric estimation of service time
  characteristics in infinite-server queues with nonstationary poisson input}.
\newblock \bibinfo{journal}{Stochastic Systems} \bibinfo{volume}{9},
  \bibinfo{pages}{183--207}.
%Type = Article
\bibitem[{Hans et~al.(2015)Hans, Chiabaut and Leclercq}]{hans2015applying}
\bibinfo{author}{Hans, E.}, \bibinfo{author}{Chiabaut, N.},
  \bibinfo{author}{Leclercq, L.}, \bibinfo{year}{2015}.
\newblock \bibinfo{title}{Applying variational theory to travel time estimation
  on urban arterials}.
\newblock \bibinfo{journal}{Transportation Research Part B: Methodological}
  \bibinfo{volume}{78}, \bibinfo{pages}{169--181}.
%Type = Article
\bibitem[{Herrera and Bayen(2010)}]{herrera2010incorporation}
\bibinfo{author}{Herrera, J.C.}, \bibinfo{author}{Bayen, A.M.},
  \bibinfo{year}{2010}.
\newblock \bibinfo{title}{Incorporation of lagrangian measurements in freeway
  traffic state estimation}.
\newblock \bibinfo{journal}{Transportation Research Part B: Methodological}
  \bibinfo{volume}{44}, \bibinfo{pages}{460--481}.
%Type = Article
\bibitem[{Jenelius and Koutsopoulos(2013)}]{jenelius2013travel}
\bibinfo{author}{Jenelius, E.}, \bibinfo{author}{Koutsopoulos, H.N.},
  \bibinfo{year}{2013}.
\newblock \bibinfo{title}{Travel time estimation for urban road networks using
  low frequency probe vehicle data}.
\newblock \bibinfo{journal}{Transportation Research Part B: Methodological}
  \bibinfo{volume}{53}, \bibinfo{pages}{64--81}.
%Type = Article
\bibitem[{Jenelius and Koutsopoulos(2015)}]{jenelius2015probe}
\bibinfo{author}{Jenelius, E.}, \bibinfo{author}{Koutsopoulos, H.N.},
  \bibinfo{year}{2015}.
\newblock \bibinfo{title}{Probe vehicle data sampled by time or space:
  Consistent travel time allocation and estimation}.
\newblock \bibinfo{journal}{Transportation Research Part B: Methodological}
  \bibinfo{volume}{71}, \bibinfo{pages}{120--137}.
%Type = Article
\bibitem[{Jin and Ma(2019)}]{jin2019non}
\bibinfo{author}{Jin, J.}, \bibinfo{author}{Ma, X.}, \bibinfo{year}{2019}.
\newblock \bibinfo{title}{A non-parametric bayesian framework for traffic-state
  estimation at signalized intersections}.
\newblock \bibinfo{journal}{Information Sciences} \bibinfo{volume}{498},
  \bibinfo{pages}{21--40}.
%Type = Book
\bibitem[{Johnson et~al.(2005)Johnson, Kemp and Kotz}]{johnson2005univariate}
\bibinfo{author}{Johnson, N.L.}, \bibinfo{author}{Kemp, A.W.},
  \bibinfo{author}{Kotz, S.}, \bibinfo{year}{2005}.
\newblock \bibinfo{title}{Univariate discrete distributions}. volume
  \bibinfo{volume}{444}.
\newblock \bibinfo{publisher}{John Wiley \& Sons}.
%Type = Techreport
\bibitem[{Kyte et~al.(2014)Kyte, Tribelhorn et~al.}]{kyte2014operation}
\bibinfo{author}{Kyte, M.}, \bibinfo{author}{Tribelhorn, M.}, et~al.,
  \bibinfo{year}{2014}.
\newblock \bibinfo{title}{Operation, analysis, and design of signalized
  intersections: a module for the introductory course in transportation
  engineering.}
\newblock \bibinfo{type}{Technical Report}. TranLIVE. University of Idaho.
%Type = Book
\bibitem[{Ni(2020)}]{ni2020signalized}
\bibinfo{author}{Ni, D.}, \bibinfo{year}{2020}.
\newblock \bibinfo{title}{Signalized Intersections}.
\newblock \bibinfo{publisher}{Springer}.
%Type = Book
\bibitem[{Prassas and Roess(2020)}]{prassas2020highway}
\bibinfo{author}{Prassas, E.S.}, \bibinfo{author}{Roess, R.P.},
  \bibinfo{year}{2020}.
\newblock \bibinfo{title}{The Highway Capacity Manual: A Conceptual and
  Research History Volume 2}.
\newblock \bibinfo{publisher}{Springer}.
%Type = Article
\bibitem[{Ramezani and Geroliminis(2012)}]{ramezani2012estimation}
\bibinfo{author}{Ramezani, M.}, \bibinfo{author}{Geroliminis, N.},
  \bibinfo{year}{2012}.
\newblock \bibinfo{title}{On the estimation of arterial route travel time
  distribution with markov chains}.
\newblock \bibinfo{journal}{Transportation Research Part B: Methodological}
  \bibinfo{volume}{46}, \bibinfo{pages}{1576--1590}.
%Type = Article
\bibitem[{Schweer and Wichelhaus(2015)}]{schweer2015nonparametric}
\bibinfo{author}{Schweer, S.}, \bibinfo{author}{Wichelhaus, C.},
  \bibinfo{year}{2015}.
\newblock \bibinfo{title}{Nonparametric estimation of the service time
  distribution in the discrete-time gi/g/∞ queue with partial information}.
\newblock \bibinfo{journal}{Stochastic Processes and their Applications}
  \bibinfo{volume}{125}, \bibinfo{pages}{233--253}.
%Type = Article
\bibitem[{Seo et~al.(2019)Seo, Kawasaki, Kusakabe and
  Asakura}]{seo2019fundamental}
\bibinfo{author}{Seo, T.}, \bibinfo{author}{Kawasaki, Y.},
  \bibinfo{author}{Kusakabe, T.}, \bibinfo{author}{Asakura, Y.},
  \bibinfo{year}{2019}.
\newblock \bibinfo{title}{Fundamental diagram estimation by using trajectories
  of probe vehicles}.
\newblock \bibinfo{journal}{Transportation Research Part B: Methodological}
  \bibinfo{volume}{122}, \bibinfo{pages}{40--56}.
%Type = Article
\bibitem[{Singh et~al.(2021)Singh, Acharya, Cruz and
  Quinino}]{singh2021estimation}
\bibinfo{author}{Singh, S.K.}, \bibinfo{author}{Acharya, S.K.},
  \bibinfo{author}{Cruz, F.R.}, \bibinfo{author}{Quinino, R.C.},
  \bibinfo{year}{2021}.
\newblock \bibinfo{title}{Estimation of traffic intensity from queue length
  data in a deterministic single server queueing system}.
\newblock \bibinfo{journal}{Journal of Computational and Applied Mathematics} ,
  \bibinfo{pages}{113693}.
%Type = Article
\bibitem[{Tan et~al.(2020)Tan, Liu, Wu, Cao and Tang}]{tan2020fuzing}
\bibinfo{author}{Tan, C.}, \bibinfo{author}{Liu, L.}, \bibinfo{author}{Wu, H.},
  \bibinfo{author}{Cao, Y.}, \bibinfo{author}{Tang, K.}, \bibinfo{year}{2020}.
\newblock \bibinfo{title}{Fuzing license plate recognition data and vehicle
  trajectory data for lane-based queue length estimation at signalized
  intersections}.
\newblock \bibinfo{journal}{Journal of Intelligent Transportation Systems} ,
  \bibinfo{pages}{1--18}.
%Type = Article
\bibitem[{Wang et~al.(2013)Wang, Bengler, Wets and Niu}]{wang2013modeling}
\bibinfo{author}{Wang, W.}, \bibinfo{author}{Bengler, K.},
  \bibinfo{author}{Wets, G.}, \bibinfo{author}{Niu, H.}, \bibinfo{year}{2013}.
\newblock \bibinfo{title}{Modeling and simulation in transportation
  engineering}.
\newblock \bibinfo{journal}{Mathematical Problems in Engineering}
  \bibinfo{volume}{2013}, \bibinfo{pages}{1--2}.
%Type = Article
\bibitem[{Wang et~al.(2020)Wang, Zhu, Ran and Jiang}]{wang2020queue}
\bibinfo{author}{Wang, Z.}, \bibinfo{author}{Zhu, L.}, \bibinfo{author}{Ran,
  B.}, \bibinfo{author}{Jiang, H.}, \bibinfo{year}{2020}.
\newblock \bibinfo{title}{Queue profile estimation at a signalized intersection
  by exploiting the spatiotemporal propagation of shockwaves}.
\newblock \bibinfo{journal}{Transportation research part B: methodological}
  \bibinfo{volume}{141}, \bibinfo{pages}{59--71}.
%Type = Article
\bibitem[{Wong et~al.(2019)Wong, Shen, Zhao and Liu}]{wong2019estimation}
\bibinfo{author}{Wong, W.}, \bibinfo{author}{Shen, S.}, \bibinfo{author}{Zhao,
  Y.}, \bibinfo{author}{Liu, H.X.}, \bibinfo{year}{2019}.
\newblock \bibinfo{title}{On the estimation of connected vehicle penetration
  rate based on single-source connected vehicle data}.
\newblock \bibinfo{journal}{Transportation Research Part B: Methodological}
  \bibinfo{volume}{126}, \bibinfo{pages}{169--191}.
%Type = Article
\bibitem[{Zhao et~al.(2021a)Zhao, Shen and Liu}]{zhao2021hidden}
\bibinfo{author}{Zhao, Y.}, \bibinfo{author}{Shen, S.}, \bibinfo{author}{Liu,
  H.X.}, \bibinfo{year}{2021}a.
\newblock \bibinfo{title}{A hidden markov model for the estimation of
  correlated queues in probe vehicle environments}.
\newblock \bibinfo{journal}{Transportation Research Part C: Emerging
  Technologies} \bibinfo{volume}{128}, \bibinfo{pages}{103128}.
%Type = Article
\bibitem[{Zhao et~al.(2021b)Zhao, Wong, Zheng and Liu}]{zhao2021maximum}
\bibinfo{author}{Zhao, Y.}, \bibinfo{author}{Wong, W.}, \bibinfo{author}{Zheng,
  J.}, \bibinfo{author}{Liu, H.X.}, \bibinfo{year}{2021}b.
\newblock \bibinfo{title}{Maximum likelihood estimation of probe vehicle
  penetration rates and queue length distributions from probe vehicle data}.
\newblock \bibinfo{journal}{IEEE Transactions on Intelligent Transportation
  Systems} .
%Type = Article
\bibitem[{Zhao et~al.(2019)Zhao, Zheng, Wong, Wang, Meng and
  Liu}]{zhao2019various}
\bibinfo{author}{Zhao, Y.}, \bibinfo{author}{Zheng, J.}, \bibinfo{author}{Wong,
  W.}, \bibinfo{author}{Wang, X.}, \bibinfo{author}{Meng, Y.},
  \bibinfo{author}{Liu, H.X.}, \bibinfo{year}{2019}.
\newblock \bibinfo{title}{Various methods for queue length and traffic volume
  estimation using probe vehicle trajectories}.
\newblock \bibinfo{journal}{Transportation Research Part C: Emerging
  Technologies} \bibinfo{volume}{107}, \bibinfo{pages}{70--91}.
%Type = Article
\bibitem[{Zheng et~al.(2018)Zheng, Jabari, Liu and Lin}]{zheng2018traffic}
\bibinfo{author}{Zheng, F.}, \bibinfo{author}{Jabari, S.E.},
  \bibinfo{author}{Liu, H.X.}, \bibinfo{author}{Lin, D.}, \bibinfo{year}{2018}.
\newblock \bibinfo{title}{Traffic state estimation using stochastic lagrangian
  dynamics}.
\newblock \bibinfo{journal}{Transportation Research Part B: Methodological}
  \bibinfo{volume}{115}, \bibinfo{pages}{143--165}.

\end{thebibliography}
\vspace{-40pt}


\end{document}